\documentstyle[twoside,12pt,epsf]{article}
\normalsize
\newcommand{\bdi}{\begin{displaymath}}
\newcommand{\edi}{\end{displaymath}}
\newcommand{\bfi}{\begin{figure}}
\newcommand{\efi}{\end{figure}}

\newcommand{\beq}{\begin{equation}}
\newcommand{\eeq}{\end{equation}}
\newcommand{\beqa}{\begin{eqnarray}}
\newcommand{\eeqa}{\end{eqnarray}}

\newcommand{\ra}{\rightarrow}
\newcommand{\Dsla}{D\hspace{-7.3pt}  /  }

\newcommand{\wt}{\widetilde}

\def\longbar#1{\setbox1=\hbox{$#1$}
\setbox2=\vbox{\hrule width 0.8\wd1}
\raise0.5\ht1\hbox{${\lower\dp1\box2}\atop\box1$}}  

\begin{document}

\begin{titlepage}

\begin{flushright}
KA--TP--25--1999 \\
\today
\end{flushright}

\vspace{1cm}
\begin{center}
{\Large \bf Multiple zero modes of the Dirac operator in three
dimensions}\\[1cm]
C. Adam* \\
Institut f\"ur Theoretische Physik, Universit\"at Karlsruhe,  \\

\medskip

\medskip

B. Muratori**,\, C. Nash*** \\
Department of Mathematical Physics, National University of Ireland, Maynooth
\vfill
{\bf Abstract} \\
\end{center}
One of the key properties of Dirac operators is the possibility of 
a degeneracy of zero modes. 
For the Abelian Dirac operator in three dimensions the construction
of multiple zero modes has been sucessfully carried out only 
very recently.
Here we generalise
these results by discussing a much wider class of Dirac operators
together with their zero modes. Further we show that those Dirac
operators that do admit zero modes may be related to Hopf maps,
where the Hopf index is related to the number of zero modes in a
simple way.   
\vfill

$^*)${\footnotesize  
email address: adam@particle.physik.uni-karlsruhe.de, adam@pap.univie.ac.at} 

$^{**})${\footnotesize
email address: bmurator@fermi1.thphys.may.ie} 

$^{***})${\footnotesize
email address: cnash@stokes2.thphys.may.ie} 
\end{titlepage}

\section{Introduction}

Fermionic zero modes of
the Dirac operator $D_A = \gamma^\mu (\partial_\mu -iA_\mu )$ are of 
importance in many places in quantum field theory and mathematical
physics \cite{AS1,JR1,JR2}. They are 
the ingredients for the computation of the index of the Dirac operator
and play a key r\^ole in understanding anomalies. In Abelian gauge 
theories, which is what we are concerned with here, they affect 
crucially the behaviour of the Fermion determinant $\det(D_A)$ 
in quantum electrodynamics. The nature  of the QED 
functional integral depends strongly on the degeneracy of the 
bound zero modes.

In three dimensions -- which is the case which we want to study here --
the first examples of such zero energy Fermion bound states were
obtained only in 1986 \cite{LoYa1}, and some further results have been found 
recently \cite{zero}. In both articles no degeneracy of these zero modes
has been observed, because, by their very methods, the authors of 
\cite{LoYa1} and of \cite{zero} could only construct one zero mode
per gauge field. Only very recently we were able to give the first
examples of Dirac operators that admit more than one zero mode
\cite{more}, thereby establishing
that the phenomenon of zero mode degeneracy exists 
for the Abelian Dirac operator in three dimensions. It is the purpose
of this article to generalise and further explain the results of
\cite{more}.

  It should be emphasised here that the problem of the existence and
degeneracy of zero modes of the
Abelian Dirac operator in three dimensions, 
in addition to being interesting in its own right, has
some deep physical implications. The authors of \cite{LoYa1} were mainly
interested in these zero modes because in an accompanying paper
\cite{FLL} it was proven that one-electron atoms with sufficiently high
nuclear charge in an external magnetic field are unstable if such zero modes 
of the Dirac operator exist. 

Further, there is an intimate connection between the existence and
number of zero modes of the Dirac operator for strong magnetic fields on
the one hand, and the nonperturbative 
behaviour of the three dimensional Fermionic
determinant (for massive Fermions) in strong external magnetic fields on the
other hand. The behaviour of this determinant, in turn, is related to
the paramagnetism of charged Fermions, see \cite{Fry1,Fry2}.
So, a thorough understanding of the zero modes of the Dirac operator
is of utmost importance for the understanding of some deep physical
problems as well.

In addition, 
it is speculated in \cite{Fry2} that the existence 
and degeneracy of zero modes  for $QED_3$ may  have a topological origin 
as it does in $QED_2$ \cite{Jac1}--\cite{Adam} --- cf. 
\cite{Fry2} for details and an account 
of the situation for $QED_{2,3,4}$. We will find some further strong
support for that conjecture in our paper.

This article is organised as follows. In Section 2 we briefly review the
case of zero modes of the Abelian Dirac operator in two dimensions,
because there exists some similarity between the general two-dimensional
case and the specific class of zero modes in three dimensions that we
want to discuss. We point out some specific features of the 
two-dimensional case that we shall need later on. In Section 3 we
review the features of maps $S^2 \ra S^2$ and of Hopf maps $S^3 \ra
S^2$, because we shall need them for a topological interpretation of
our results. In Section 4 we construct our class of Dirac operators 
together with their zero modes.
Further we show that the corresponding
magnetic fields may be related to Hopf maps (they may be expressed
as Hopf curvatures of some Hopf maps), and that the Hopf index is
related to the number of zero modes of a given Dirac operator in a
simple fashion. This topological interpretation of the magnetic fields
requires the introduction of a fixed, universal background magnetic
field. In the final section we briefly describe another class of 
multiple zero modes that were not covered in the main section. Further
we discuss how our results are related to a rigorous upper bound on
the growth of the number of zero modes for strong magnetic fields, and
we provide some interpretations for the fixed, universal background
field that we had to introduce.

\section{Two-dimensional case}

First of all, we want to briefly recall the situation in two dimensions,
because there will be some analogies with the class of three-dimensional
zero modes that we shall discuss below. The two-dimensional Dirac equation is
\beq
\gamma_\mu (-i\partial_\mu -A_\mu (x))\Psi (x) =0,
\eeq
where $x= (x_1 ,x_2)$, $\mu =1,2$, $\gamma_\mu =\sigma_\mu$ and $\Psi$
is a two-component spinor. Both in Euclidean space ${\rm\bf R}^2$ and on the
two-sphere $S^2$ all zero modes are either left-handed (i.e., the
lower component of $\Psi$ is zero) or right-handed (the upper component of
$\Psi$ is zero). Further, a solution of the first type (left-handed) may be 
mapped into a solution of the second type by the simple replacement
$A_\mu \ra -A_\mu$, therefore we may restrict to the left-handed case
\beq
-i \left( \begin{array}{cc} 0 & \partial_z -iA_z  \\ \partial_{\bar z}
-iA_{\bar z} & 0 \end{array} \right) \rho^{1/2}(z,\bar z)e^{i\lambda (z,
\bar z)}  \left( \begin{array}{c} 1  \\ 0 \end{array} \right) =0
\eeq
where 
\beq
z=x_1 +ix_2 \, , \quad \partial_z =\frac{1}{2}(\partial_1 -i\partial_2 )
\eeq
\beq
A_z =\frac{1}{2}(A_1 -iA_2) .
\eeq
Here $\rho^{1/2}(z,\bar z)$ is a real, nonsingular function and $\lambda$ is
a pure gauge factor that has to be determined accordingly (see below).

At this point we want to make some observations. Firstly, obviously only the 
left lower component $\partial_{\bar z} -iA_{\bar z}$ of the Dirac operator
acts on the spinor in (2). Therefore, a spinor that solves (2) may be
multiplied by an arbitrary holomorphic function $f(z)$  and still solves the
same Dirac equation (2). A more complicated way of stating the same
observation (which will be useful for the three-dimensional case) is as
follows. We search for a function $f(z, \bar z)$ such that
\beq
-i\gamma_\mu (\partial_\mu f) \left( \begin{array}{c} 1 \\ 0 
\end{array} \right) =0,
\eeq
then $f\Psi$ will formally solve the Dirac equation for the same Dirac 
operator (i.e., the same gauge potential) as $\Psi$. A possible choice
for $f$ is $f=z$ and, as a consequence of the Leibnitz rule, arbitrary
functions $f(z)$ of $z$ only are allowed. Observe that (5) implies
\beq
\det (-i\gamma_\mu \partial_\mu f)=(f_{,1})^2 + (f_{,2})^2 =0,
\eeq
which requires a complex $f$.

Secondly, from eq. (2) $A_\mu$ may be expressed in terms of $\rho$ and
$\lambda$ as ($\epsilon_{12}=1$)
\beq
A_\mu =\frac{1}{2}\epsilon_{\mu\nu}\partial_\nu \ln \rho +\lambda_{,\mu}
\eeq
\beq
A_z =\frac{i}{2}\partial_z \ln \rho +\lambda _{,z} .
\eeq
Now assume that $\rho (z,\bar z)$ has a zero at some point $z_0$. As $\rho$ 
is real, let us assume that the zero is 
of the type $((z-z_0)(\bar z-\bar z_0))^\alpha$
for some $\alpha >0$. This zero induces a singular contribution
$A_\mu^{\rm sing}$ to the gauge potential $A_\mu$
(here $z_0 =: y_1 +iy_2$), where
\beq
A_\mu^{\rm sing}=\alpha \epsilon_{\mu\nu}\frac{x_\nu -y_\nu}{(\vec x - \vec y
)^2} = \alpha \partial_\mu {\rm arg} (z-z_0) .
\eeq 
From the r.h.s. of (9) it is obvious that $A_\mu^{\rm sing}$ is in fact
a pure gauge. Therefore, all singularities of $A_\mu$ due to zeros
of $\rho$ of the above type
may be gauged away by choosing the appropriate gauge functions
\beq
\lambda =-\alpha \, {\rm arg}(z-z_0) := \alpha \, \arctan\frac{x_2 -y_2}{
x_1 -y_1}
\eeq
in (2). As we want nonsingular gauge potentials, this gauge choice will be 
assumed in the sequel. However, $\lambda$ in (10) is not a single-valued
function, and the gauge factor $\exp (i\lambda)$ of the spinor in (2)
will be single-valued only provided that $\alpha =n \in {\rm\bf N}$, i.e.,
only zeros of the above type of integer order are allowed for $\rho$.
Zeros of other types (as e.g. $\rho =z+\bar z$, which is zero at $z=0$)
lead to singularities in $A_\mu$ that are not pure gauge, i.e., they lead to
singular magnetic fields (14). They are, therefore, forbidden.

Now suppose that a zero mode for a non-singular gauge field is given and
$\rho$ has some zeros of integer order of the allowed type
as just described. For each zero
$((z-z_0)(\bar z-\bar z_0))^n$ we may multiply the zero mode in (2) by
the function $f(z) = (z-z_0)^{-n}$. This is a function of $z$ only, 
therefore the new spinor $f\Psi$ is a zero mode of the same
Dirac operator. As a consequence, for each Dirac operator that admits
zero modes there exists a zero mode (2) such that $\rho$ is
nonzero everywhere,  $\rho^{1/2}$ is strictly positive, 
$\rho^{1/2}>0$. Further, the corresponding pure gauge terms (10) are
absent, and we may assume that the gauge factor in (2) is absent
altogether, $\lambda \equiv 0$, which corresponds to Lorentz gauge
$\partial_\mu A_\mu =0$ for the gauge potential in (7).

Let us assume that a spinor (2) is given with $\lambda =0$, $\rho^{1/2}>0$
and 
\beq
\lim_{|z|\to\infty} \rho \sim (z\bar z)^{-\alpha_\infty} .
\eeq
Square-integrability of $\Psi$ in ${\rm\bf R}^2$ implies $\alpha_\infty
>1$. If $n+1 >\alpha_\infty >n \, , \, n \in {\rm\bf N}$, 
then further square-integrable
zero modes of the same Dirac operator may be constructed as
\beq
\Psi_k = z^k\rho^{1/2}
 \left( \begin{array}{c} 1  \\ 0 \end{array} \right) 
 \, ,\quad k=0 \ldots n-1
\eeq
and the above-mentioned zero modes with some zeros of integer order may be 
recovered as linear combinations of (12). In addition, $\alpha_\infty$
determines the magnetic flux $\Phi$,
\beq
\Phi =\int d^2 xB =2\pi \alpha_\infty
\eeq
\beq
B=\partial_1 A_2 - \partial_2 A_1 =2\partial_z \partial_{\bar z} \ln\rho .
\eeq
Here we may follow two different approaches. Either we assume that our results
really exist in Euclidean space. Then there are no further restrictions on 
$\alpha_\infty$. Further, whenever $\alpha_\infty =n\in {\rm\bf N}$, 
then there are only $n-1$ square-integrable zero modes (12), because
the one with $k=n-1$ is not square-integrable (its $L^2$ norm behaves as
$\ln V$, where $V$ is the volume of space).

Or, on  the other hand, we could interpret $z$ as a stereographic 
coordinate on 
the Riemann sphere. Then $z=\infty$ is a single point, and $\rho$ in (11) has 
a zero at this point if $\alpha_\infty \neq 0$, which leads to a singularity
in $A_\mu$. This singularity, however, cannot be removed by a gauge
transformation without introducing a singularity somewhere else. 
Instead, two different gauge potentials have to be chosen on different
coordinate patches (e.g. on the northern and southern hemisphere), such that
the difference of the two gauge potentials in the overlap region is a
pure gauge $\partial_\mu \lambda$. The gauge function $\exp (i\lambda )$
(which acts on the zero mode) is single-valued only if $\alpha_\infty
\equiv n_\infty \in {\rm\bf N}$ and, consequently, the magnetic flux
$\Phi =2\pi n_\infty$ is quantised (in fact, this is just the well-known
topology of the Dirac monopole). In addition, the zero modes are now 
normalised w.r.t. the integration measure on the sphere, therefore there are
$n_\infty =\Phi/(2\pi)$ normalisable zero modes, in accordance with the 
index theorem.

\section{Maps $S^2 \ra S^2$ and Hopf maps $S^3 \ra S^2$}

The second homotopy group of the two-sphere is nontrivial, $\Pi_2 (S^2)=
{\bf Z}$, therefore
maps $S^2 \ra S^2$ are characterised by the integer 
winding number $w$. One way 
of describing them is by interpreting both $S^2$ as Riemann spheres
and by introducing stereographic coordinates $z\in {\rm\bf C}$ on both
of them. A specific class of such maps $S^2 \ra S^2$ may then be described
by rational maps
\beq
R: \, z\ra R(z) =\frac{P(z)}{Q(z)}
\eeq
where $P(z)$ and $Q(z)$ are polynomials, and $z$ and $R(z)$ are interpreted as 
stereographic coordinates on the domain and target $S^2$, respectively. The
winding number $w$ of this map is given by the degree of the map,
\beq
w={\rm deg}(R)={\rm max}(p,q)
\eeq
where $p$ and $q$ are the degrees of the polynomials $P(z)$ and $Q(z)$
\cite{HMS,Horv}.
Another possibility of computing the same winding number involves the
pullback under $R(z)$ of the standard area two-form $\Omega$ on $S^2$
(in stereographic coordinates),
\beq
\Omega =\frac{2}{i}\frac{d\bar z dz}{(1+z\bar z)^2} \, , \quad
\int \Omega =4\pi .
\eeq
The pullback is ($'$ means derivative w.r.t. the argument)
\beq
R^* \Omega =\frac{2}{i}\frac{|R' (z)|^2}{(1+R\bar R)^2}d\bar z dz
\eeq
and obeys
\beq
\int R^* \Omega =4\pi w
\eeq
where $w$ is again the winding number (16). 

However, rational maps are not the
only types of functions that generate maps $S^2 \ra S^2$. Instead of 
$R(z)$ we may e.g. choose the functions (here $z=u^{1/2}\exp (i\varphi )$,
$u:=z\bar z$ and $f$ is an at the moment arbitrary real function)
\beq
G(z,\bar z)=f(u)z^n =: g^{1/2}(u)e^{in\varphi}
\eeq
which we shall need later on. The pullback of $\Omega$ under $G$ is
($'\equiv \partial_u$)
\beq
G^* \Omega =2n\frac{g'}{(1+g)^2}dud\varphi
\eeq
and its integral is
\beq
\int G^* \Omega =4\pi n\int_0^\infty du\frac{g'}{(1+g)^2}=
-4\pi n\frac{1}{1+g(u)}|_0^\infty .
\eeq
If $g(0) =0$ and $g(\infty) =\infty$, as holds e.g. for the rational maps
$R(z)=z^n$, then the function $G$ in (20) defines a map $S^2 \ra S^2$ with
winding number $n$,
\beq
\int G^* \Omega =4\pi n .
\eeq
Apart from $g(u) \ge 0$, which follows from the definition of $g$, $g$ is not 
very much restricted in the intermediate range $0<u<\infty$. 
Let us, e.g., assume that $g$
has a singularity at $u_1$ and a zero at $u_2$ (we assume $u_1 < u_2$
for this example), then the region $u\in [0,u_1]$ of the domain $S^2$ is
mapped onto the target $S^2$ with winding number $+n$, the region $[u_1 ,
u_2]$ is mapped onto the target $S^2$ with winding number $-n$, and the 
region $[u_2 ,\infty]$ is mapped onto the target $S^2$
with winding number $+n$, again. Therefore the net
winding number is $n$.

Observe that it is possible to relate the pullback $R^* \Omega$ or
$G^* \Omega$ to a magnetic field $B$ via (e.g. for $G$)
\beq
G^* \Omega =:Bdx_1 dx_2 \equiv F ,
\eeq
where $F=(1/2)F_{\mu\nu}dx_\mu dx_\nu $ is the magnetic field strength
two-form.
However, all $B$'s that are constructed in this way have an even integer 
multiple of $2\pi$ as magnetic flux, $\Phi =\int d^2 x B =4\pi n=2\pi \cdot
2n$, as is obvious from (23). Differently stated, if we want to formally
express magnetic fields with magnetic fluxes that are odd integer
multiples of $2\pi$ 
by maps $R$ or $G$, then we have to allow for square-root type,
double-valued maps $R\sim z^{n/2}$ or $G\sim z^{n/2}$. This we shall need
later on.

 Hopf maps are maps $S^3 \ra S^2$. 
The third homptopy group of the two-sphere is non-trivial as well,
$\Pi_3 (S^2) ={\bf Z}$, therefore such maps are characterised by an
integer topological index, the so-called Hopf index.
Hopf maps may be expressed, e.g., by maps $\chi :
{\rm\bf R}^3 \ra {\rm\bf C}$ provided that the complex function $\chi$
obeys $\lim_{|\vec x|\ra \infty} \chi (\vec x)=\chi_0 ={\rm const}$,
where $\vec x =(x_1 ,x_2 ,x_3)^{\rm T}$. The pre-images in ${\rm\bf R}^3$
of points of the target $S^2$ (i.e., the pre-images of points $\chi =
{\rm const}$) are closed curves in ${\rm\bf R}^3$ (circles in the 
related domain $S^3$). Any two different circles are linked $N$ times,
where $N$ is the Hopf index of the given Hopf map $\chi$.
Further, a magnetic field $\vec {\cal B}$ 
(the Hopf curvature) is related to the
Hopf map $\chi$ via
\beq
\vec {\cal B} = 
\frac{2}{i}\frac{(\vec\partial\bar\chi)\times(\vec\partial\chi)}{
(1+\bar\chi \chi)^2} =2\frac{(\vec\partial T)\times\vec
\partial\sigma}{(1+T)^2}
\eeq
where $\chi =Se^{i\sigma}$ is expressed in terms of its modulus $S=:T^{1/2}$
and phase $\sigma$ at the r.h.s. of (25).

Mathematically, the curvature ${\cal F}=\frac{1}{2}{\cal F}_{ij}dx_i dx_j$, 
${\cal F}_{ij}= \epsilon_{ijk}{\cal B}_k$, 
is the pullback under the Hopf map, ${\cal F}=\chi^* \Omega$,
of the standard area two-form $\Omega$ , (17), on the target $S^2$. 
Geometrically, $\vec {\cal B}$ is 
tangent to the closed curves $\chi ={\rm const}$ (see e.g. 
\cite{Ran1,FN1,BS1,JaPi3}; the authors of \cite{JaPi3}
describe Hopf curvatures slightly
differently, by the Abelian projection of an $SU(2)$ pure gauge
connection, which has some technical advantages).
The Hopf index $N$ of $\chi$ may be computed from $\vec {\cal B}$ via
\beq
N=\frac{1}{16\pi^2}\int d^3 x \vec {\cal A} \vec {\cal B}
\eeq
where $\vec {\cal B}=\vec\partial\times \vec {\cal A}$.

Once a Hopf map $\chi$ is given, we may construct further Hopf maps by 
composing the Hopf map $\chi$ with maps $S^2 \ra S^2$,
\beq
\chi_G :S^3 \stackrel{\chi}{\ra} S^2 \stackrel{G}{\ra}S^2
\eeq
where $G$ might be e.g. a $G(\chi ,\bar\chi)$ as in (20) or a rational map
$R(\chi)$ as in (15). Further, if $\chi$ has Hopf index $N=1$ and $G$ has
degree (i.e. winding number) $n$, then the composed Hopf map $\chi_G$
has Hopf index $N=n^2$.

The simplest (standard) Hopf map $\chi$ with Hopf index $N=1$ is
\beq
\chi =\frac{2(x_1 +ix_2)}{2x_3 -i(1-r^2)}
\eeq
with modulus and phase
\beq
T:= \bar\chi \chi = \frac{4(r^2 -x_3^2)}{4x_3^2 +(1-r^2)^2}\, ,\quad
\sigma =\arctan\, \frac{x_2}{x_1} + \arctan\, \frac{1-r^2}{2x_3} 
\eeq
\beq
\sigma =\sigma^{(1)} + \sigma^{(2)} \, ,\quad \sigma^{(1)}= \arctan
\frac{x_2}{x_1} \, ,\quad \sigma^{(2)} =\arctan\frac{1-r^2}{2x_3} .
\eeq
Here the phase $\sigma$ is a sum of two terms $\sigma^{(1)}$ and 
$\sigma^{(2)}$, where $\sigma^{(1)}$ is multiply valued around the singular
point $\chi =0$ in target space, i.e., along the $x_3$ axis in the domain
${\rm\bf R}^3$, and $\sigma^{(2)}$ is multiply valued around the singular
point $\chi =\infty$, i.e., around the circle $\{ \vec x \in {\rm\bf R}^3
\, \backslash \, x_3 =0, x_1^2 +x_2^2 =1 \}$.
As $\chi$ in three dimensions will play a role similar to $z=x_1 +ix_2$ in
two dimensions in Section 2, it is important to note a crucial difference
in this respect. The same phase $\varphi ={\rm arg}\, z =\arctan(x_2 /
x_1)$ is multiply valued around both singular points $z=0$ and $z=\infty$
in the two-dimensional case.

The simplest standard Hopf map, (28), leads to the Hopf curvature 
\beq
\vec {\cal B} = \frac{16}{(1+r^2)^2} \vec N
\eeq
and we have introduced the unit vector
\beq
\vec N = \frac{1}{1+r^2}
\left( \begin{array}{c} 2x_1 x_3 -2x_2  \\ 2x_2 x_3 +2x_1 \\
1-x_1^2 -x^2_2 +x_3^2 \end{array} \right) 
\eeq
($\vec N^2 =1$) for later convenience.

\section{Three-dimensional case} 

Here we want to study multiple 
solutions of the three-dimensional, Abelian Dirac equation
(the Pauli equation)
\beq
-i\sigma_i \partial_i \Psi (x) 
=A_i (x) \sigma_i \Psi (x),
\eeq
where $\vec x =(x_1 ,x_2 ,x_3)^{\rm T}$, $i,j,k = 1\ldots 3$, $\Psi$ is a
two-component, square-integrable spinor on ${\rm\bf R}^3$, $\sigma_i$
are the Pauli matrices and $A_i$ is an Abelian gauge potential. 
Before starting the actual computations we want to mention some general
aspects of the Dirac equation (33). Firstly, for any pair $(\Psi ,\vec A)$
that solves the Dirac equation (33), the zero mode $\Psi$ has to obey
\beq
\vec \partial \vec \Sigma =0
\eeq
where $\vec\Sigma$ is the spin density of $\Psi$,
\beq
\vec \Sigma =\Psi^\dagger \vec \sigma \Psi .
\eeq
Secondly, when a spinor $\Psi$ is given that obeys (34) (i.e., it is a possible
zero mode), then the corresponding gauge potential $\vec A$ that solves
the Dirac equation (33) together with $\Psi$ may actually be expressed in
terms of $\Psi$ \cite{LoYa1},
\bdi
A_i =\frac{1}{|\vec \Sigma |}(\frac{1}{2}\epsilon_{ijk}\partial_j
\Sigma_k +{\rm Im}\, \Psi^\dagger \partial_i \Psi )   
\edi
\beq 
= \frac{1}{2}\epsilon_{ijk}(\partial_j \ln |\vec \Sigma |){\cal N}_k
+\frac{1}{2}\epsilon_{ijk}\partial_j {\cal N}_k +
{\rm Im}\, \widehat\Psi^\dagger
\partial_i \widehat\Psi
\eeq
where we have expressed $ A_i$ in terms of the general
unit vector and unit spinor
\beq
\vec {\cal N}=\frac{\vec \Sigma}{|\vec \Sigma |}\, ,\quad \widehat\Psi =
\frac{\Psi}{|\Psi^\dagger \Psi |^{1/2}} 
\eeq 
for later convenience.

Next we want to discuss the simplest example of a zero mode that was already
found in \cite{LoYa1}, because we need it as a starting point. The
authors of \cite{LoYa1} observed that a solution to this equation could
be obtained from a solution to the simpler equation
\beq
-i\vec\sigma \vec\partial \Psi =h\Psi
\eeq
for some scalar function $h(x)$. In this case the corresponding gauge field
that obeys the Dirac equation (33) together with the spinor (38) is given
by
\beq
A_i =h \frac{\Psi^\dagger \sigma_i \Psi}{\Psi^\dagger \Psi} .
\eeq
In addition, they gave the following explicit example
\beq
\Psi =\frac{4}{(1+r^2)^{\frac{3}{2}}}({\bf 1} +i\vec x \vec \sigma )
\left( \begin{array}{c} 1  \\ 0 \end{array} \right) 
\eeq
\beq
\vec \Sigma =\Psi^\dagger \vec \sigma \Psi =\frac{16}{(1+r^2)^2}\vec N
\eeq
where $\vec N$ is the specific unit vector defined in (32) and we chose
the factor 4 in (40) for later convenience.
The spinor (40) obeys
\beq
-i \vec\sigma \vec\partial  \Psi = \frac{3}{1+r^2}\Psi
\eeq
and is, therefore, a zero mode for the gauge field
\beq
\vec A = \frac{3}{1+r^2} \frac{\Psi^\dagger \vec\sigma \Psi}{\Psi^\dagger \Psi}
=\frac{3}{1+r^2} \vec N
\eeq
with magnetic field $\vec B =\vec\partial\times \vec A$ 
\beq
\vec B 
=\frac{12}{(1+r^2)^2} \vec N 
\eeq
($\vec N$ is the unit vector defined in (32)).

Now we want to repeat the argument of (5) and (6) of the two-dimensional
case, i.e., we assume that a function $\chi$ exists such that
\beq
(-i\sigma_j \partial_j \chi )({\bf 1} +i\vec x \vec \sigma )
\left( \begin{array}{c} 1  \\ 0 \end{array} \right)  =0.
\eeq
Consequently, 
$\chi^n \Psi$, $n\in {\rm \bf Z}$ (where $\Psi$ is the zero mode (40)),
are additional formal zero modes for the same gauge field (43).
Condition (45) implies
\beq
{\rm det} (-i\vec\sigma\vec\partial\chi ) =\sum_{i=1}^3 \chi_{,i}\chi_{,i}
=0,
\eeq
therefore, $\chi$ necessarily must be complex. Indeed, such a function
$\chi$ fulfilling (45) exists. It is just the simplest Hopf map $\chi$, (28), 
as may be checked easily. For the formal zero modes $\chi^n
\Psi$ we observe the following two points. Firstly, $n$ has to be 
integer, because only integer powers of $\chi$ lead to a single-valued
spinor $\chi^n \Psi$. Secondly, $\chi^n \Psi$ is singular for all
$n\in {\rm\bf Z}\setminus \{ 0\}$, because $\chi$ is singular along the circle
$\{ \vec x \in {\rm\bf R}^3 \, \backslash \, x_3 = 0 \, 
, x_1^2 + x_2^2 =1\}$ and
zero along the $x_3$ axis. Therefore, the formal zero modes $\chi^n \Psi$,
with $\Psi$ given in (40), are not acceptable. However, we shall find some
zero modes, different from (40), where multiplication with $\chi^n$ will
lead to acceptable new zero modes for some $n\neq 0$. 

For this purpose, let us observe that the spin density (41) of the simplest
zero mode (40) is in fact equal to the Hopf curvature (31) of the 
simplest Hopf 
map (28) (we chose the constant factor 4 in (40) in order to achieve this
equality; otherwise $\vec \Sigma$ would only be proportional to the Hopf
curvature, which is enough for our purposes). As a consequence (see eq. (25))
\beq
\vec \Sigma^{(M)}:=e^{M(\chi ,\bar\chi )}\vec \Sigma =
\frac{16}{(1+r^2)^2}e^{M(\chi ,\bar\chi )}\vec N
\eeq
still is the spin density of a zero mode, i.e., it still obeys $\vec
\partial \vec \Sigma^{(M)} =0$. Here $M(\chi ,\bar\chi)$ is a real function
of $\chi$ and $\bar\chi$. The corresponding zero mode reads
\beq
\Psi^{(M)} =e^{i\Lambda}e^{M/2}\Psi = e^{i\Lambda}e^{M/2}
\frac{{\bf 1}+i\vec\sigma \vec x}{(1+r^2)^{3/2}}
\left( \begin{array}{c} 1  \\ 0 \end{array} \right) 
\eeq
where $\Lambda$ is a gauge function that has to be determined accordingly
(analogously to our discussion in Section 2; see below). $\Psi^{(M)}$
is proportional to the simplest zero mode (40), therefore it remains true
that additional formal zero modes for the same Dirac operator may
be constructed from $\Psi^{(M)}$ by multiplication with powers $\chi^n$
of $\chi$, (28). 

At this point we want to present some first examples of such multiple
zero modes (these examples were already discussed in \cite{more}).
For this purpose, we need some more results of \cite{LoYa1}. 
The authors of \cite{LoYa1} observed that, in addition to their simplest 
solution (40), they could find similar solutions to eq. (38) with higher 
angular momentum. Using instead of the constant spinor $(1,0)^{\rm T}$
the spinor
\beq
\Phi_{l,m}= \left( \begin{array}{c} \sqrt{l+m+1/2}\, Y_{l,m-1/2} \\
 -\sqrt{l-m+1/2}\, Y_{l,m+1/2} \end{array} \right)
\eeq
(where $m\in [-l-1/2\, ,\, l+1/2]$ and $Y$ are spherical harmonics), 
they found the solutions
\beq
\Psi_{l,m}=r^l (1+r^2)^{-l-\frac{3}{2}}({\rm\bf 1}+
i\vec x \vec \sigma )\Phi_{l,m}
\eeq
\beq
\vec A_{l,m}=(2l+3)(1+r^2)^{-1}\frac{\Psi^\dagger_{l,m}\vec\sigma
\Psi_{l,m}}{\Psi^\dagger_{l,m} \Psi_{l,m}} .
\eeq
Specifically, for maximal magnetic quantum number $m=l+1/2$, these solutions
read
\beq
\Psi_l :=\Psi_{l,l+1/2} = \frac{Y_{l,l} r^l}{(1+r^2)^{l+3/2}}
({\rm\bf 1}+ i\vec x \vec \sigma )
\left( \begin{array}{c} 1  \\ 0 \end{array} \right) 
\eeq
\beq
\vec A^{(l)} 
=\frac{3+2l}{1+r^2} \vec N
\eeq
\beq
\vec B^{(l)}  
=\frac{4(3+2l)}{(1+r^2)^2} \vec N 
\eeq
(where we have omitted an irrelevant constant factor in (52)).
Hence, $\Psi_l$ is proportional to the simplest zero mode (40) and is, 
therefore, still an eigenvector of the matrix $-i\sigma_j \partial_j \chi$
with eigenvalue zero. Further, 
the zero mode $\Psi_l$ may be rewritten as (again, we ignore irrelevant 
constant factors)
\beq
\Psi_l = e^{il\varphi}\Bigl( \frac{T}{1+T}\Bigr)^{l/2}
(1+r^2)^{-3/2}
({\rm\bf 1}+ i\vec x \vec \sigma )
\left( \begin{array}{c} 1  \\ 0 \end{array} \right) 
\eeq
where we introduced polar coordinates $(x_1 ,x_2 ,x_3) \ra (r,\theta ,
\varphi)$,
$T$ is the squared modulus (29), and (up to an irrelevant constant) 
\beq
Y_{l,l} = e^{il\varphi}\sin^l \theta =e^{il\varphi}\frac{(r^2 -
x_3^2)^{l/2}}{r^l}
= e^{il\varphi}\frac{(1+r^2)^l}{r^l}\Bigl( \frac{T}{1+T}\Bigr)^{l/2} .
\eeq
Taking further into account that $\varphi =\arctan (x_2 / x_1)
=\sigma^{(1)}$ we conclude that the spinors
\beq
\Psi_{n,l}=\chi^{-n}\Psi_l = e^{i(l-n)\sigma^{(1)} -in\sigma^{(2)}}
\frac{T^{(l-n)/2}}{(1+T)^{l/2}} 
\Psi \, , \quad n=0,\ldots l
\eeq
are non-singular, square-integrable zero modes for the same gauge field
$\vec A^{(l)}$ and, therefore, the Dirac operator with gauge field
$\vec A^{(l)}$ given by (53) has $l+1$ square-integrable zero modes (57).
Here $\sigma^{(1)}$ and $\sigma^{(2)}$ are the two terms (30) of the
phase of the simplest Hopf map (28).

At this point several remarks are necessary. Firstly, observe that the
function $\exp (M)$, as defined in (47), for the zero modes (57) reads
\beq
e^M = \frac{T^{l-n}}{(1+T)^l}
\eeq
\beq
\lim_{T\to 0}e^M \sim T^{l-n} \, ,\quad \lim_{T\to \infty}e^M \sim T^{-n}
\eeq
Hence, $\exp (M)$ has a zero of order $l-n$ at $T=0$ and a zero of order 
$n$ at $T=\infty$. As in the two-dimensional case, these zeros introduce
singularities in the gauge potentials, which are cured by the pure gauge 
functions $(l-n)\sigma^{(1)}$ and $-n\sigma^{(2)}$, respectively, leading
to the well-behaving gauge potentials (53). In contrast to the two-dimensional
case, the singularity at $\chi =\infty$ may be cured independently, i.e.,
without introducing singularities somewhere else (for an explanation see 
below).

Secondly, we observe that already the simplest magnetic field (44) (for $l=0$)
is proportional but not equal to the Hopf curvature (31) (the magnetic field
has a factor of 12 instead of 16, i.e., they differ by $4(1+r^2)^2 \vec N$).
Here we will take the following point of view. We assume that this difference 
is related to a fixed, universal background magnetic field $\vec B^{\rm b}$,
\beq
\vec B^{\rm b}=-\frac{4}{(1+r^2)^2}\vec N
\eeq
which couples to the Fermion via the Dirac operator but is ``non-dynamical''
otherwise. Then for the ``dynamical'' part $\wt B_j:= B_j - B_j^{\rm b}$ of
$B_j$ it holds that
\beq
\wt B_j = B_j - B_j^{\rm b}=\frac{16}{(1+r^2)^2}N_j ={\cal B}_j
\eeq
where ${\cal B}_j$ is the Hopf curvature (31). We immediately find that
this feature continues to hold for all the higher $B_j^{(l)}$ in (54),
\beq
 \wt B_j^{(l)} = B_j^{(l)} - B_j^{\rm b}=\frac{16(1+l/2)}{(1+r^2)^2}N_j .
\eeq
These $\wt B_j^{(l)}$ are Hopf curvatures for the Hopf maps
\beq
\chi^{(l)}=T^{1/2}e^{i(1+l/2)\sigma}
\eeq
where $T$ and $\sigma$ are given in (29). We find that we have to allow
for double-valued, square-root type Hopf maps if we want to relate all
$\wt B_j^{(l)}$ to Hopf curvatures. Further, we find the relation
\beq
N=\Bigl( \frac{k+1}{2}\Bigr)^2
\eeq
between the Hopf index $N$ and the number $k=l+1$ of zero modes. We will 
find that after the subtraction of the universal background field (60)
all these features continue to hold for a much wider class of solutions
to the Dirac equation. 

In order to discuss this wider class, let us go back to the general zero
mode (48) which depends on a function $M(\chi ,\bar\chi)$ and a pure
gauge function $\Lambda$. The corresponding gauge potential $\vec 
A^{(M)}$ that obeys the Dirac equation together with $\Psi^{(M)}$ may
be computed from (36),
\beq
A^{(M)}_j = A_j +\frac{1}{2}\epsilon_{jkl}M_{,k}N_l + \Lambda_{,j}
\eeq
\beq = A_j +\frac{i}{2}(M_{,\chi}\chi_{,j} - M_{,\bar\chi}\bar\chi_{,j})
+ \Lambda_{,j}
\eeq 
where the second line follows after some algebra.  Here $A_j$ is the gauge 
potential (43) of the simplest zero mode (40) and $N_l$ is the unit vector
(32).

At this point we have to discuss the possible singularities of $\vec A^{(M)}$,
which will determine $\Lambda$ and, at the same time, pose some restrictions
on $\exp (M)$, as in the two-dimensional case (Section 2). As in the
two-dimensional case, zeros of $\exp (M)$ cause singularities of 
$\vec A^{(M)}$, and in order to cause only removable singularities, these 
zeros have to be of the type (with possible multiplicity $n$)
\beq
((\chi -z_i)(\bar\chi - \bar z_i))^n =:\zeta^n \bar\zeta^n
\eeq
which implies for the above expression (66) (without the pure gauge piece
$\Lambda_{,j}$) 
\beq
\frac{i}{2}(M_{,\chi}\chi_{,l}
- M_{,\bar\chi}\bar\chi_{,l}) \sim \frac{in}{2}\frac{\bar \zeta
\chi_{,l} - \zeta \bar\chi_{,l}}{\zeta \bar\zeta} + \, \ldots
\eeq
where the remainder is regular at $\zeta =0$. The above singularity may be
compensated by the pure gauge factor
\beq
\Lambda = -n \, \arctan \frac{i(\zeta - \bar\zeta)}{\zeta + \bar\zeta} .
\eeq
Indeed ($\zeta_{,l} \equiv \chi_{,l}$),
\beq
\Lambda_{,l} = -\frac{in}{2}\frac{\bar \zeta
\zeta_{,l} - \zeta \bar\zeta_{,l}}{\zeta \bar\zeta}
\eeq
precisely cancels the singular term (68). The spinor in (48) 
is multiplied by the
gauge factor $\exp (i\Lambda)$. This factor is single-valued only if the
order $n$ of the zero  is integer, because $\Lambda$ 
in (69) 
is a multiply-valued function.

In fact, this is not yet the whole story about singularities in 
$\bar A^{(M)}_l$. The point is that the expression
\beq
\frac{i}{2}\frac{\bar \chi \chi_{,l} - \chi \bar\chi_{,l}}{\chi\bar\chi}
\eeq
is singular in the limit $\chi \to \infty$ as well, as may be easily
checked. 
Further, the derivatives of the gauge factors, (70), for all the
zeros  (67) produce this expression (71) for $\chi\to\infty$,
because $\lim_{\chi \to \infty}\zeta =\chi$. In addition, $\exp (M)$ may
cause a similar term (71) for $\vec A^{(M)}$ if it behaves as
\beq
\lim_{|\chi |\to \infty}\exp (M) \sim |\chi \bar\chi |^{-n_\infty}
\equiv T^{-n_\infty} .
\eeq
Here $n_\infty$ has to be a positive integer or zero, as we shall see
immediately. Further, $\exp (M)$ has to reach the limiting value
sufficiently fast,
\beq
\lim_{|\chi | \to \infty} 
(T^{n_\infty} \exp (M)) \sim 1+cT^{-\alpha} \, ,\quad \alpha \ge 1
\eeq
($c$ is some constant) as will be explained below.
Therefore, altogether we have to compensate
\beq
(-n_\infty + \sum_i n_i) \frac{i}{2}\frac{\bar \chi \chi_{,l} - 
\chi \bar\chi_{,l}}{\chi\bar\chi}
\eeq
by
an additional gauge transformation, without introducing further singularities
at $\chi =0$ (here $n_i$ are the multiplicities of the zeros $z_i$ of
$\exp (M)$). 

Fortunately this is possible for the following reason. If we were to 
compensate (74) by the full gauge function
\beq
\Lambda = (n_\infty - \sum_i n_i)\arctan 
\frac{i(\chi - \bar\chi)}{\chi + \bar\chi}
=-(n_\infty - \sum_i n_i) \sigma
\eeq
(where $\sigma$ is the phase of $\chi$ given in (29)), this would introduce 
a singularity at $\chi =0$. However, $\sigma$ is the sum of two terms
$\sigma = \sigma^{(1)} + \sigma^{(2)}$ (see (30)),
where $\sigma^{(1)}_{,l}$ is singular at $\chi =0$ and $\sigma^{(2)}_{,l}$
is singular at $\chi = \infty$. Therefore, we may cancel the singularity
of (74) without introducing further singularities by
performing an additional gauge transformation using only $\sigma^{(2)}$,
\beq
\Lambda = -(n_\infty - \sum_i n_i) \sigma^{(2)} .
\eeq  
Obviously, $n_\infty$ has to be integer for (76) to be an acceptable gauge
function.

We want to emphasise again here that there is a crucial difference to the
two-dimensional case (Section 2, last paragraph), where no gauge choice was 
possible to achieve a non-singular gauge potential for all $z$. The reason
for this difference lies in the different topological features of the
underlying spaces $S^2$ and $S^3$, respectively. In fact, 
the second cohomology group of the $S^2$ is non-trivial, $H_2 (S^2) =
{\bf Z}$. Therefore, 
it is not possible to find a globally defined gauge potential
on $S^2$ for magnetic fields with non-zero (quantised) magnetic flux.
On the other hand, $H_2 (S^3) =0$, therefore it is always possible to find a
well-behaving non-singular gauge potential for a well-behaving non-singular
magnetic field.

One consequence of the above discussion is that (as in the two-dimensional
case) the zeros $((\chi -z_0)(\bar\chi -\bar z_0))^n$ may be removed by
multiplying the corresponding zero mode with the holomorphic function
(in the variable $\chi$) $(\chi -z_0)^{-n}$ without changing the Dirac 
operator. Therefore, for each Dirac operator that admits zero modes 
there exists one zero mode such that $\exp (M/2)$ is strictly positive,
$\exp (M/2) > 0$ for all $\chi <\infty$. This we will assume in the sequel.
Further we assume
\beq
\lim_{|\chi| \to\infty} \exp (M) \sim T^{-n_\infty}
\eeq
as in (72), (73). The corresponding zero mode is 
\beq
\Psi^{(M)} = e^{i\Lambda}e^{M/2}\Psi
\eeq
where $\Lambda$ is given in (76) (with $n_i =0$). Additional non-singular,
square-integrable zero modes for the same Dirac operator are
\beq
\Psi^{(M)}_n =\chi^n \Psi^{(M)} \, ,\quad n=0, \ldots n_\infty
\eeq
i.e., there are $k=n_\infty +1$ zero modes. As in the two-dimensional case,
zero modes with arbitrary allowed zeros may be constructed as linear 
combinations of the above zero modes (79).

Finally we have to discuss the related magnetic field.
The magnetic field $ B^{(M)}_i = \epsilon_{ijk}
\partial_j  A^{(M)}_k$ corresponding to the gauge potential (65) is
\bdi
B^{(M)}_l = B_l +\frac{1}{2}[M_{,\chi} 
(\chi_{,lk}N_k +\chi_{,l}N_{k,k}
-\chi_{,kk}N_l - \chi_{,k}N_{l,k})+
\edi
\bdi
M_{,\bar\chi} 
(\bar\chi_{,lk}N_k +\bar\chi_{,l}N_{k,k}
-\bar\chi_{,kk}N_l - \bar\chi_{,k}N_{l,k})
\edi
\beq
-(M_{,\chi\chi}\chi_{,k}\chi_{,k} +M_{,\bar\chi \bar\chi}\bar\chi_{,k}
\bar\chi_{,k} +2M_{,\chi \bar\chi}\chi_{,k}\bar\chi_{,k}) N_l]
\eeq
where $ B_l$ is the magnetic field (44). 
After some tedious algebra we find that only the coefficient of
$M_{,\chi \bar\chi}$ is nonzero, i.e.,
\beq
\chi_{,lk}N_k +\chi_{,l}N_{k,k}
-\chi_{,kk}N_l - \chi_{,k}N_{l,k} =0
\eeq
\beq
\chi_{,k}\chi_{,k} =0
\eeq
\beq
\chi_{,k}\bar\chi_{,k} = 8\frac{(1+\chi\bar\chi)^2}{(1+r^2)^2}
\eeq
and, therefore
\beq
B^{(M)}_l =  B_l -8\frac{(1+\chi\bar\chi)^2}{(1+r^2)^2}
M_{,\chi\bar\chi}N_l .
\eeq
Obviously, $\vec B^{(M)}$ will be finite in the limit $|\chi | \to \infty$
only if $\lim_{|\chi | \to\infty}M_{,\chi \bar\chi} \sim 
|\chi\bar\chi |^{-2-\epsilon} \, , \, \epsilon \ge 0$. This corresponds to
eq. (73) and explains our remark that $\exp (M)$ has to reach its limiting
value sufficiently fast.

As in (61), we now have to subtract the background magnetic field (60) in 
order to be able to relate the resulting ``dynamical'' magnetic field
$\wt B_l^{(M)}$ to Hopf maps. We find
\beq
\wt B_l^{(M)} =\Bigl( 1-\frac{1}{2}(1+\chi \bar\chi )^2 M_{,\chi \bar\chi}
\Bigr) {\cal B}_l
\eeq
where $\vec {\cal B}$ is the Hopf curvature (31).

At this point we want to specialise to the class of functions
\beq
M(\chi ,\bar\chi )=M(\chi\bar\chi)\equiv M(T)
\, , \quad M' \le 0
\eeq
($'\equiv \partial_T$)
 because we want to relate
them to Hopf maps of the type (27) where the function $G$ is given by (20).
For these functions $M(T)$, (85) simplifies to
\beq
\wt B_l^{(M)} =\Bigl( 1-\frac{1}{2}(1+T )^2 (M' +TM'' )
\Bigr) {\cal B}_l .
\eeq
We want to re-express this magnetic field as a Hopf curvature $\vec 
{\cal B}^{(G)}$ for the Hopf map
\beq
\chi^{(G)} =g^{1/2}(T)e^{im\sigma}
\eeq
which is a composition of the standard Hopf map (28) and a map $S^2 \ra
S^2$ of the type $G$ as in (20). The Hopf curvature $\vec {\cal B}^{(G)}$ is
\beq
\vec {\cal B}^{(G)}=2m\frac{(\vec\partial g)\times \vec\partial \sigma}{
(1+g)^2}=m\frac{g'(1+T)^2}{(1+g)^2}\vec {\cal B}
\eeq
which is indeed a Hopf curvature if $g(0)=0 \, , \, g(\infty)=\infty$, 
see (22).
Equality of (87) and (89) implies
\beq
-m \Bigl( \frac{1}{1+g}\Bigr) ' = -\Bigl( \frac{1}{1+T}\Bigr) '
-\frac{1}{2}(M' T)'
\eeq
or upon integration
\beq
\frac{m}{1+g}=\frac{1}{1+T} + \frac{1}{2}TM' +\frac{1}{2}n_\infty
\eeq
\beq
m=1+\frac{1}{2}n_\infty
\eeq
(where we have chosen an appropriate constant of integration in (91)).
Here $M' \le 0$ (together with $\exp (M)>0$ and condition (77))
is a sufficient condition to ensure $g\ge 0$.

Therefore, we find that for all the zero modes of the type (78), (86) the
corresponding magnetic fields may indeed be expressed as Hopf curvatures,
provided that we allow for double-valued Hopf maps, $m=1+(n_\infty /2)$,
whenever the Dirac operator has an even number of zero modes. In addition,
we confirm the general relation between Hopf index $N=m^2$ and the number
of zero modes $k=n_\infty +1$,
\beq
N=\Bigl( \frac{k+1}{2}\Bigr)^2 .
\eeq

\section{Discussion}

We have found a class of magnetic fields (87) that are the Hopf curvatures of
the Hopf maps (88) (after the subtraction of the fixed background field
(60)). The corresponding Dirac operator shows a degeneracy of zero modes,
where the number of zero modes is related to the Hopf index via eq. (93).
Here we had to allow for double-valued Hopf maps whenever the number of
zero modes is even.

Further, we imposed some restrictions on the zero modes (i.e., on the
functions $M(\chi ,\bar\chi)$) because we wanted to relate the
corresponding magnetic fields to the specific, simple type (88) of
Hopf maps. We think that these restrictions are a mere technicality,
and that abandoning these restrictions  will just lead to more
complicated Hopf maps. One specific type of such Hopf maps, different
from (88), is easily accessible and leads to results that are in complete
agreement with the ones we have described above, therefore we want to
 describe it briefly.

Recall that there exists a class of Hopf maps that are a composition
of the standard Hopf map with an arbitrary rational map $R(\chi) =
P(\chi)/Q(\chi)$, see (15). The corresponding Hopf curvature reads
($'\equiv$ derivative w.r.t. the argument)
\beq
{\cal B}_l^{(R)}=\frac{| P' Q -PQ' |^2}{(| P |^2 + |Q|^2)^2}
(1+\chi\bar\chi)^2 {\cal B}_l =\wt B_l^{(M_R)}
\eeq
($P$ and $Q$ do not have a common zero), where we have already indicated on
the r.h.s. of (94) that there exists a magnetic field $\wt B_l^{(M_R)}$
for some zero mode $\Psi^{(M_R)}$. In fact, $\exp (M_R)$ reads
\beq
\exp (M_R) = \frac{(1+\chi\bar\chi)^2}{(|P|^2 + |Q|^2)^2} 
\eeq
\beq
\lim_{|\chi|\to \infty}\exp (M_R) =(\chi\bar\chi)^{-2(w-1)}
\eeq
where $w$ is the degree (16) of the rational map $R$. Therefore there are
$k=2w-1$ zero modes
\beq
\Psi_n^{(M_R)}=\chi^n \Psi^{(M_R)} \, , \quad n=0, \ldots 2(w-1).
\eeq
In addition, the corresponding magnetic field $\wt B_l^{(M_R)}$ (after
the subtraction of the background field) is indeed equal to the Hopf
curvature (94), as may be computed easily with the help of eq. (84).
The Hopf index is $N=w^2$, therefore the relation (93) between Hopf
index and number of zero modes is confirmed once more.

This class of solutions has another interesting feature. A zero mode
may be constructed (a specific linear combination of the zero modes (97))
such that its spin density $\Sigma_l$ equals the magnetic field $\wt
B_l^{(M_R)}$. Hence in addition to the Dirac equation (33) this 
solution obeys the equation $\Sigma_l =\wt B_l$. This system of
equations of motion is generated by the Lagrangian density
\beq
{\cal L} = \Psi^\dagger \sigma_j (-i\partial_j -A_j)\Psi + \frac{1}{2}
\wt A_j \wt B_j ,
\eeq 
where the background field is coupled to the Fermion,
but it is absent in the second, ``kinetic'' term (the Abelian
Chern--Simons term). This explains why we called $\wt A_l$ the
``dynamical'' gauge potential (for details see \cite{hoin,holi}, where
these solutions (``Hopf instantons'') were discussed in depth).

Another point that we want to mention here is the fact that our results
may be used to estimate the number of zero energy bound states (zero
modes) for strong magnetic fields. This is seen especially easily for
the higher angular momentum zero modes (57), because  the
magnetic fields (54) for higher angular momentum $l$ are just multiples 
of the simplest magnetic
field (44). Therefore, the number $k=l+1$ of zero modes  for strong 
magnetic fields (i.e. large $l$) behaves like
\beq
k = l+1 \sim c\int d^3 x |\vec B^{(l)}|
\eeq
(it holds that $\lim_{|\vec x| \to \infty}|\vec B^{(l)}| \sim r^{-4}$, 
therefore the integral in (21) exists),
i.e., $k$ grows linearly with the strength of the magnetic field (here $c$
is some constant). This remains true in a certain sense for our other
solutions. From (93) we infer that the number of zero modes $k$ behaves
like $k\sim N^{1/2}$ for large $k$ ($N$ is the Hopf index). Further, as
$N \sim \int d^3 x \wt A_j \wt B_j$, the number of zero modes grows
like $\lambda$ under a rescaling $\wt A_j \ra \lambda \wt A_j$, 
$\wt B_j \ra \lambda \wt B_j$. 
  This is well within the rigorous upper bound on the
possible growth of the number of zero modes
\beq
k\sim c' \int d^3 x |\vec B|^{3/2}
\eeq
that was first stated in \cite{LoYa1} and later derived in \cite{Fry2} 
(here $c'$ is a constant; the difference between $\wt B_j$ and $B_j$ is
unimportant for strong fields, because the background magnetic field (60)
is the same for all magnetic fields). We should mention here that it is, in
principle, possible that the Dirac operators of 
our magnetic fields have in fact more zero
modes than we have discovered with our methods, which would imply that
the true number of zero modes is closer to the rigorous upper bound (100).

Observe that it was possible to  relate our magnetic fields to Hopf curvatures
only after the subtraction of the fixed, universal background field (60)
(although the existence and degeneracy of the zero modes per se does not 
require the background field). Further, the above-mentioned solutions
to the equations of motion of the Chern--Simons and Fermion system (98)
only exist in the presence of this background field, as well 
(\cite{hoin,holi}). Therefore, this background field (60) seems to be
rather fundamental for our discussion, and one wonders whether it
admits some further interpretation. We cannot yet give a final answer
to this question, but we want to mention two possible interpretations
that were already given in \cite{hoin}. On one hand, if one compares
the background magnetic field (60) with the magnetic fields (54) of the
higher angular momentum zero modes (52), then one realises that 
changing the angular momentum by one unit produces a change of the
corresponding magnetic field that is precisely minus two times the
background field (60). It is, therefore, tempting to conjecture that
the background field is somehow related to the half-integer angular
momentum (spin) of the Fermion.
 Of course, this is just an observation at this point, because a mechanism
that generates this background field is still missing.

On the other hand,
it is possible to re-interpret the background gauge potential 
$\vec A^{\rm b} = -(1+r^2)^{-1} \vec N$ of the background magnetic field
(60) as a spin connection $\omega$ in the Dirac equation (33) on a
conformally flat manifold with torsion. Generally, the Dirac operator with
spin connection reads (see e.g. \cite{Bertl} for details)
\beq
\Dsla = \gamma^a E_a{}^\mu (\partial_\mu +A_\mu +\frac{1}{4}[\gamma_b ,\gamma_c
]\omega^{bc}{}_\mu )
\eeq
where $\gamma^a$ ($\equiv \sigma^a$ in our case) are the usual Dirac matrices,
$E_a{}^\mu$ is the inverse vielbein and $\omega^{bc}{}_\mu$ is the spin
connection (here $\mu ,\nu$ are Einstein (i.e., space time) indices and
$a,b,c$ are Lorentz indices). Our Dirac equation (33) may be rewritten 
in the form of eq.  
(101) provided that the vielbein is conformally flat, $E_a{}^\mu =f\delta_a^\mu
$, where $f$ is an arbitrary function. Using $[\sigma_b ,\sigma_c ]=
2i\epsilon_{bcd}\sigma^d$ we find
\beq
\frac{i}{2}\delta_a^k \epsilon_{bcd}\sigma^a \sigma^d \omega^{bc}{}_k
\stackrel{!}{=} \delta_a^k \sigma^a A^{\rm b}_k
\eeq
(here $k$ is an Einstein index in three dimensions). The l.h.s. of (102) has
to be antisymmmetric in $a,d$, i.e., the quantity $\wt \omega_{da}
:= \delta_a^k \epsilon_{bcd}\omega^{bc}{}_k$ obeys $\wt \omega_{da}
=-\wt\omega_{ad}$. This leads to $\wt \omega_{ab} =\epsilon_{abc}
\delta_c^k A^{\rm b}_k$. If we further assume $\omega^{ab}{}_k =
-\omega^{ba}{}_k$ (i.e., covariant constancy of the metric) then we find that
\beq
\omega_{abk} =\delta_{ka}A^{\rm b}_b - \delta_{kb}A^{\rm b}_a
\eeq
(where $A^{\rm b}_a \equiv \delta_a^k A^{\rm b}_k$, i.e., it is {\em not}
the Lorentz vector $E_a{}^k A^{\rm b}_k$). Finally, we find for the torsion
$T$ (expressed in Lorentz indices only)
\beq
2T_{abc}= (\delta_{ab}\delta_c^k - \delta_{ac}\delta_b^k )\partial_k f
- (\omega_{abc} - \omega_{acb})
\eeq
where
\beq
\omega_{abc}=E_c{}^k\omega_{abk}=f\delta_c^k \omega_{abk} .
\eeq
Hence, with $\omega_{abk}$ given by (103), we may freely choose a conformally 
flat metric (i.e., conformal factor $f$) and compute the resulting torsion 
via (104). Due to the form of $\omega_{abk}$ (i.e., $\vec A^{\rm b}$) it is,
however, not possible to choose a conformal factor such that the torsion is
zero. On the other hand, it is possible to choose the flat metric $f=1$,
so that (the anti-symmetric part of) the spin connection is given just by 
the torsion. 
      
Finally we want to point out that some important questions still remain
to be answered. 

Firstly, all our zero modes are of a specific type. They are multiples
(by a scalar function) of the simplest spinor (40). There exist, of
course, zero modes of a different type (see e.g. \cite{zero}). 
By the very methods of \cite{zero}, only one zero mode per Dirac
operator (i.e., per gauge potential) could be constructed. We believe
that the methods of this paper may, in principle, be adapted to 
address the question of a degeneracy of zero modes for more general
Dirac operators, like those in \cite{zero}.

Secondly, all our magnetic fields are Hopf curvatures after the
subtraction of the background magnetic field (60), where one has to
allow for double-valued Hopf maps in the case of an even number of
zero modes. This immediately leads to the question whether this
feature can be proven in general, and whether the existence and
degeneracy of zero modes may be explained on topological grounds,
as is the case in even dimensions.

Thirdly, the topological interpretation of our magnetic fields
(as Hopf curvatures) was possible only after the introduction of
the universal background magnetic field (60). We already provided
some possible interpretations of this background field, but we
think that it plays a rather fundamental role in the whole problem
and, therefore, deserves some further investigation.

 Anyhow, we think that our results should be relevant for some future
developments
in mathematical physics, as well as for the understanding of 
non-perturbative aspects of quantum
electrodynamics, especially in three dimensions.

\section{Acknowledgments}
The authors thank M. Fry for helpful discussions. In addition,
CA gratefully acknowledges useful conversations with R. Jackiw.
CA was supported by a Forbairt Basic Research Grant during part of the
work.
BM gratefully acknowledges financial support from the Training and 
Mobility of Researchers scheme (TMR no. ERBFMBICT983476).

\end{document}